\documentstyle[preprint,revtex]{aps}
\begin{document}
\preprint{FERMILAB-PUB-92/264--T}
\preprint{LBL-32987}
\preprint{hep-ph@xxx/yymmnn}
\begin{title}
Flavor Asymmetry of the Nucleon Sea:\\
Consequences for Dilepton Production
\end{title}
\author{Estia J. Eichten\thanks{Internet address: eichten@fnal.fnal.gov.}}
\begin{instit}
Fermi National Accelerator Laboratory
P.O. Box 500, Batavia, Illinois 60510
\end{instit}
\moreauthors{Ian Hinchliffe\thanks{Internet address:
hinchliffe\%theorm.hepnet@lbl.gov.}}
\begin{instit}
Lawrence Berkeley Laboratory
Berkeley, California 94720
\end{instit}
\moreauthors{Chris Quigg\thanks{Internet address: quigg@fnal.fnal.gov.}}
\begin{instit}
Fermi National Accelerator Laboratory
P.O. Box 500, Batavia, Illinois 60510
\end{instit}
\begin{abstract}
Parton distributions derived from a chiral quark model that generates an excess
of down quarks and antiquarks in the proton's sea satisfactorily describe the
measured yields of muon pairs produced in proton-nucleus collisions.
Comparison of
dilepton yields from hydrogen and deuterium targets promises greater
sensitivity
to the predicted flavor asymmetry.
\end{abstract}
In a recent article \cite{ehq}, we showed that the fluctuation of
constituent quarks into quarks plus Goldstone bosons generates a
flavor-asymmetry in the light-quark sea of the nucleon that is a
plausible origin for the violation of the Gottfried sum rule reported by
the New Muon Collaboration (NMC) at CERN \cite{nmc}.  Fermilab experiment E772
has now presented measurements of the yields of massive muon pairs in
collisions of 800-GeV$/c$ protons with nuclear targets that are
sensitive to the flavor content of the nucleon sea \cite{e772}.  We show
in this paper that our picture of the Gottfried-sum-rule defect
also accounts for the new dilepton data, and we make predictions for a
more sensitive test using hydrogen and deuterium targets.

Forward (Feynman-$x_F \;{\raisebox{-.4ex}{\rlap{$\sim$}}
\raisebox{.4ex}{$>$}}\;0.1$)
production of massive dilepton pairs in high-energy
collisions of protons with nuclear targets is dominated by the annihilation of
a
$u$-quark from the beam with a $\bar{u}$-antiquark from the target, and so is
sensitive to the distribution of antiquarks in the target nucleons.
Charge symmetry relates the distribution of up-antiquarks in the neutron to
the distribution of down-antiquarks in the proton,
\begin{equation}
\bar{u}^{(n)}(x,Q^2) = \bar{d}^{(p)}(x,Q^2)  \;\;\; ,
\end{equation}
where $x$ is Bjorken's scaling variable and $Q^2$ labels the scale on which the
parton distributions are measured.
The yield per nucleon $\sigma_A$ in proton collisions
with nucleus $A$ differs from the yield per nucleon $\sigma_{\rm isoscalar}$ in
proton collisions with an isoscalar target by the factor
\begin{equation}
R_A(x) \equiv \frac{\sigma_A(x)}{\sigma_{\rm isoscalar}(x)} \approx
    1 + \frac{(N-Z)}{A} \; \frac{\bar{d}(x)-\bar{u}(x)}
    {\bar{d}(x)+\bar{u}(x)} \;\;\; ,\label{rats}
\end{equation}
where $A$, $Z$, and $N$ are the atomic weight, atomic number, and number of
neutrons in the target, and the antiquark densities refer to the proton.
The approximate equality follows upon neglect of $d\bar{d}$ annihilations.
The deviation of $R_A(x)$ from unity measures the
flavor-asymmetry of the light-quark sea.

The possibility that the light-quark sea contains unequal numbers of up
and down quarks has been raised by the New Muon Collaboration's
determination \cite{nmc} of the integral
\begin{equation}
I_G = \int_{0}^{1} dx\;\frac{\left[F_2^{\mu p}(x,
Q^2) - F_2^{\mu n}(x,Q^2)\right]}{x} = 0.240 \pm 0.016\;\;\; .
\label{nmcig}
\end{equation}
In the quark-parton model, the integral can be expressed as
\begin{equation}
I_G = \frac{1}{3}+\frac{2}{3}\int_{0}^{1} dx
\left[\overline{u}(x,Q^2) - \overline{d}(x, Q^2)\right]~~~.\label{3}
\end{equation}
The Gottfried sum rule \cite{gott}, $I_G = 1/3$, follows from the
assumption that the sea is up-down symmetric.  The observed defect
implies a small excess,
\begin{equation}
\int_{0}^{1} dx \left[\overline{u}(x,Q^2) - \overline{d}(x, Q^2)\right]
= -0.14 \pm 0.024 \;\;\; ,
\end{equation}
of down quarks in the sea.

Because we expect $I_G$ to be essentially independent of $Q^2$ \cite{sach}, it
is
convenient to analyze the flavor content of the sea at a momentum scale
relevant to hadron structure, where the important degrees of freedom are
constituent quarks, Goldstone bosons, and gluons.  Consider a proton
composed of three
constituent quarks: $uud$.  An excess of down quarks over up quarks in
the sea arises naturally from the
isospin-respecting fluctuation of the constituent quarks into quarks and
pions, the lightest of the Goldstone bosons, {\em viz.}
$u\rightarrow (\pi^+d, \pi^0u)$ and $d\rightarrow
(\pi^0d, \pi^-u)$.  If $a$ denotes the probability for a constituent up
quark to turn into a down quark and a $\pi^+$ (containing a $u$-quark
and a $\bar{d}$-antiquark), the proton composition after one iteration
is $(2+7a/4)u+(1+11a/4)d+(7a/4)\overline{u}+(11a/4)\overline{d}$.  The
valence composition remains  $u_v=(u-\overline u) = 2$ and
$d_v=(d-\overline d) = 1$, but the sea contains an excess of down quarks
and antiquarks over up quarks and antiquarks.

In Ref. \cite{ehq}, we implemented this picture in the
framework of the effective chiral quark model formulated by Manohar and
Georgi \cite{GM}.  A straightforward calculation of the probability for
a constituent quark to fluctuate leads to $I_G = 0.278$, encouragingly
close to the experimental value (\ref{nmcig}).  After adjusting the
ultraviolet cutoff of the chiral quark model to better reproduce the
observed Gottfried-sum-rule defect, we constructed parton distributions
based on the Eichten-Hinchliffe-Lane-Quigg (EHLQ) Set~1 distributions
\cite{ehlq}.  These new distributions give a good account of the NMC
measurements of the Gottfried integral, the difference
$F_2^{\mu p}-F_2^{\mu n}$, and the ratio $F_2^{\mu n}/F_2^{\mu p}$.
Details of the construction of the parton distributions and the
comparison with data may be found in Ref. \cite{ehq}.  For present
purposes, it is important to note that the flavor-asymmetric sea generated by
chiral field theory, which is concentrated at small values of $x$, is a
small perturbation on the flavor-symmetric sea of EHLQ Set 1
\cite{ehlq},  $x\bar{u}(x) = x\bar{d}(x) = 0.182(1-x)^{8.54}$.

The production of massive muon pairs in proton-nucleus collisions
offers another window on the composition of the sea.  Fermilab
experiment E772 has compared yields from the isoscalar targets
$^2$H and C with yields from a neutron-rich target, W.  According to
Eq. (\ref{rats}), the ratio of yields per nucleon \cite{moss} is
$R_W(x) \approx 1+0.195(\bar{d}(x)-\bar{u}(x))/(\bar{d}(x)+\bar{u}(x))$.
Because the measured ratio shown in Figure \ref{e772f1} is consistent
with unity, the authors of Ref. \cite{e772} concluded that there is no
evidence for a large flavor-asymmetry in the light-quark sea of the
nucleon.  There is, however, no conflict between the dilepton results
and the Gottfried-sum-rule defect observed by NMC.

The chiral field theory calculation reviewed above leads with no readjustment
of parameters to the thick solid
curve \cite{details} shown in Figure \ref{e772f1}.  That prediction is entirely
consistent with the E772 data.  So, too, is the flavor-asymmetric fit (``D0'')
made to deeply inelastic lepton scattering data by Martin, Roberts,
and Stirling \cite{mrs}.  The effect on $R_W$ is small because the
ratio of $\bar{u}(x)/\bar{d}(x)$ is everywhere close to unity for both
the chiral quark model and the MRS(D0) fit.

The data do discriminate against {\em ad hoc} modifications of the EHLQ
structure functions considered by Ellis and Stirling \cite{ellis} and by
us \cite{ehq}.  The dashed line in Figure \ref{e772f1} shows the
prediction of the modified EHLQ structure functions with
$\bar{u}(x)/\bar{d}(x) = (1-x)^{5.4}$ that we examined in
Ref. \cite{ehq}.  This form magnifies the effect of a flavor asymmetry
upon $R_W$ at large values of $x$, whereas the Gottfried integral is
determined by the magnitude of the $(\bar{d}-\bar{u})$ excess, which is
determined at small $x$.  Even in the interval $0.04\,<\,x\,<\,0.15$,
however, the {\em ad hoc} form predicts a larger asymmetry than is
observed.

Ellis and Stirling \cite{ellis} proposed to examine the shape near
$x_F=0$ of the differential cross section $m^3d\sigma/dx_Fdm$ for the
reaction $pd\rightarrow \mu^+\mu^-+$ anything as a probe of the
difference between the proton and neutron sea distributions.  We show
in Figure \ref{e772f2} the Drell-Yan cross section predicted using the
structure functions obtained from the chiral quark model.  The
lowest-order calculation has been normalized to the large-$x_F$ data
using a $K$-factor of 1.4.  It gives an excellent fit to the E772 data,
without the suppression of the $x_F\,<\,0$ cross section given by the
{\em ad hoc} parton distributions of Ellis and Stirling (cf. Figure 2
of Ref. \cite{e772}).

Direct comparison of the yield of
dileptons from hydrogen and deuterium targets maximizes the sensitivity
of the ratio $R_A$ to a flavor asymmetry, because
$R_p \approx 1 - (\bar{d}(x)-\bar{u}(x))/(\bar{d}(x)+\bar{u}(x))$.  A new
experiment has been proposed using the
E772 apparatus to make this measurement \cite{p866}.  The prediction of
the chiral quark model, shown as the thick solid curve in Figure \ref{p866f1},
is slightly below unity because hydrogen is a (maximally)
neutron-{\em poor}
nucleus.  The effect of the flavor asymmetry in the nucleon sea is again
small, as it is for the MRS(D0) structure functions plotted as the
dotted curve.  As expected, the {\em ad hoc} modification of the EHLQ
structure functions produces a very large effect.  Future experiments
should be able to discriminate against this extreme possibility.  Making
the case for or against the flavor-asymmetric sea predicted by the chiral quark
model presents a considerable
challenge to dilepton experiments.

Dilepton production in hadron-nucleus collisions complements deeply
inelastic lepton scattering as a probe of the composition of the
light-quark sea of the nucleon.  Lepton scattering is the more sensitive
probe at small $x$, because the difference $F_2^{\mu p}-F_2^{\mu n}$ is
determined by the difference $\bar{d}(x)-\bar{u}(x)$, whereas the Drell-Yan
process has greater sensitivity at large $x$, where the parton densities
are small, because it measures the fractional difference
$(\bar{d}(x)-\bar{u}(x))/(\bar{d}(x)+\bar{u}(x))$.  We find the chiral quark
model
mechanism compelling and we look forward to new experimental tests of
the small excess of $\bar{d}$ over $\bar{u}$ it implies at low values of $x$.

We thank Keith Ellis for supplying predictions for the MRS(D0)
structure functions.  Fermilab is operated by Universities Research
Association, Inc., under
contract DE-AC02-76CHO3000 with the United States Department of Energy.
This work was supported at LBL by the Director, Office of Energy
Research, Office of High Energy and Nuclear Physics, Division of High
Energy Physics of the U. S. Department of Energy under Contract
DE-AC03-76SFO0098.

\figure{The ratio $R_W\equiv \sigma_W/\sigma_{\rm isoscalar}$ of
dilepton yields per nucleon from tungsten and isoscalar targets as a
function of $x_{\rm target}$.  The data are from Fermilab experiment
E772, Ref. \cite{e772}.  Open circles at small $x$ are the
ratio before correction for nuclear shadowing.  The thick solid curve is our
prediction based on chiral field theory, using the full Drell-Yan cross section
with parton distributions evaluated at the dimuon mass.  The thin solid curve
is calculated using the parton distributions from chiral field theory at fixed
$Q^2=5~({\rm GeV}/c)^2$ in Eq. (\ref{rats}).  The dotted curve shows the
prediction of the MRS(D0) parton distributions \cite{mrs}, using
Eq. (\ref{rats}).  An {\em ad hoc} modification of the EHLQ Set 1
structure functions (Ref. \cite{ehlq}) yields the dashed curve (Drell-Yan cross
section, parton distributions evolved to $Q^2=m^2$)  and the dot-dashed curve
(Eq. (\ref{rats}), fixed
$Q^2=5~({\rm GeV}/c)^2$).
\label{e772f1}}
\figure{Differential cross section $m^3d\sigma/dx_Fdm$ as a function of
$x_F$ for the reaction $pd \rightarrow \mu^+\mu^- +{\rm anything}$ at
800 GeV from Fermilab experiment E772, Ref. \cite{e772}.  The solid
curve is our prediction based on chiral field theory for a dimuon mass
$m=8.15~{\rm GeV}/c^2$. \label{e772f2}}
\figure{The ratio $R_p\equiv \sigma_p/\sigma_{d}$ of dilepton yields per
nucleon from hydrogen and deuterium targets as a
function of $x_{\rm target}$.  The thick solid curve is our prediction based on
chiral field theory, using the full Drell-Yan cross section with parton
distributions evaluated at the dimuon mass.  The thin solid curve is calculated
using the parton distributions from chiral field theory at fixed $Q^2=5~({\rm
GeV}/c)^2$ in Eq. (\ref{rats}).  The dotted curve shows the
prediction of the MRS(D0) parton distributions \cite{mrs}, using
Eq. (\ref{rats}).  An {\em ad hoc} modification of the EHLQ Set 1
structure functions (Ref. \cite{ehlq}) yields the dashed curve (Drell-Yan cross
section, parton distributions evolved to $Q^2=m^2$)  and the dot-dashed curve
(Eq. (\ref{rats}), fixed
$Q^2=5~({\rm GeV}/c)^2$).\label{p866f1}}

\begin{references}
\bibitem[9]{details}We have evolved the chiral quark model distributions
to values of $Q^2$ corresponding to the E772 data points and evaluated
the Drell-Yan cross section.  The thin solid line results from neglecting the
effects of $Q^2$-evolution and of
$d\bar{d}$ annihilations.  Those effects are seen to be are small.
\end{references}
\end{document}